\documentclass[12pt]{article}

\usepackage{amsmath}
\usepackage{amssymb}
\usepackage{amstext}
\usepackage{amsopn}
\usepackage{amsfonts}
\usepackage{amsxtra}
\usepackage[english]{babel}
\usepackage{graphicx}
\usepackage{float}
\usepackage{bm}
\usepackage{multirow}
\usepackage{hyperref}
\usepackage{lineno}
\usepackage{color}
\usepackage[superscript]{cite}
\usepackage{times}
\usepackage{fullpage}

\makeatletter
\renewcommand\@biblabel[1]{#1.}
\makeatother

\newcommand{\icm}{\ensuremath{\textrm{cm}^{-1}}}

\newcommand{\blue}{\textcolor{blue}}

\newcommand{\LNO}{La$_{3}$Ni$_{2}$O$_{7}$}
\newcommand{\EF}{$E_{\text{F}}$}

\begin{document}


\begin{trivlist}
  \item[] {\Large\textsf{\textbf{Electronic correlations and partial gap in the bilayer nickelate La$_{3}$Ni$_{2}$O$_{7}$}}}
\end{trivlist}


\begin{trivlist}
  \item[] Zhe~Liu$^{1,\ast}$, Mengwu~Huo$^{2,\ast}$, Jie~Li$^{1,\ast}$, Qing~Li$^{1}$, Yuecong Liu$^{1}$, Yaomin~Dai$^{1,\dag}$, Xiaoxiang~Zhou$^{1}$, Jiahao~Hao$^{1}$, Yi~Lu$^{1,\dag}$, Meng~Wang$^{2,\dag}$, \& Hai-Hu~Wen$^{1,\dag}$


 \item[] $^{1}$\emph{National Laboratory of Solid State Microstructures and Department of Physics, Collaborative Innovation Center of Advanced Microstructures, Nanjing University, Nanjing 210093, China}
 \item[] $^{2}$\emph{Center for Neutron Science and Technology, Guangdong Provincial Key Laboratory of Magnetoelectric Physics and Devices, School of Physics, Sun Yat-Sen University, Guangzhou, Guangdong 510275, China}


  \vspace{2mm}
  \item[] $^{\ast}$These authors contributed equally to this work.
  \item[] $^{\dag}$email: ymdai@nju.edu.cn; yilu@nju.edu.cn; wangmeng5@mail.sysu.edu.cn; hhwen@nju.edu.cn
  \vspace{4mm}
\end{trivlist}


\boldmath
\begin{trivlist}
\item[] {\bf The discovery of superconductivity with a critical temperature of about 80~K in La$_{3}$Ni$_{2}$O$_{7}$ single crystals under pressure has received enormous attention. La$_{3}$Ni$_{2}$O$_{7}$ is not superconducting under ambient pressure but exhibits a transition at $T^{\ast} \simeq 115$~K. Understanding the electronic correlations and charge dynamics is an important step towards the origin of superconductivity and other instabilities. Here, our optical study shows that La$_{3}$Ni$_{2}$O$_{7}$ features strong electronic correlations which significantly reduce the electron's kinetic energy and place this system in the proximity of the Mott phase. The low-frequency optical conductivity reveals two Drude components arising from multiple bands at the Fermi level. The transition at $T^{\ast}$ removes the Drude component exhibiting non-Fermi liquid behavior, whereas the one with Fermi-liquid behavior is barely affected. These observations in combination with theoretical results suggest that the Fermi surface dominated by the Ni-$d_{3z^{2}-r^{2}}$ orbital is removed due to the transition at $T^{\ast}$. Our experimental results provide pivotal information for understanding the transition at $T^{\ast}$ and superconductivity in La$_{3}$Ni$_{2}$O$_{7}$.}
\end{trivlist}
\unboldmath

%

%
Since the discovery of superconductivity with a transition temperature $T_{c} \simeq$ 9--15~K in the thin films of hole-doped infinite-layer nickelates Nd$_{1-x}$Sr$_{x}$NiO$_{2}$~\cite{Li2019Nature,Li2020PRL,Zeng2020PRL}, tremendous efforts have been made to find more superconducting nickelates and raise their $T_{c}$. To date, several other doped rare-earth nickelates such as (La/Pr)$_{1-x}$Sr$_{x}$NiO$_{2}$~\cite{Osada2020NL,Osada2021AM,Wang2022NC,Sun2023AM} and La$_{1-x}$Ca$_{x}$NiO$_{2}$~\cite{Zeng2022SA}, and the stoichiometric quintuple-layer Nd$_{6}$Ni$_{5}$O$_{12}$~\cite{Pan2022NM} have been found to exhibit superconductivity with $T_{c}$ up to 18.8~K~\cite{Sun2023AM}. Applying pressure can enhance the onset superconducting temperature of Pr$_{0.82}$Sr$_{0.18}$NiO$_{2}$ monotonically from 17~K at ambient pressure to 31~K at 12.1~GPa without showing any trend towards saturation~\cite{Wang2022NC}. Despite the above progress achieved in thin films, the search for evidence of superconductivity in bulk materials seems extraordinarily challenging~\cite{Li2020CM,Wang2020PRM}.

Recently, trace of superconductivity with a $T_{c}$ = 80~K was observed in bulk single crystals of the bilayer Ruddlesden-Popper (R-P) phase \LNO\ under high pressure~\cite{Sun2023Nature}, arousing a flurry of excitement in the community of high-$T_{c}$ superconductivity~\cite{Hou2023arXiv,Zhang2023arXivYuan,Wang2024PRX,Luo2023PRL,Zhang2023PRB,Yang2023arXiv,Lechermann2023arXiv,Gu2023arXiv,Shen2023arXiv,Sakakibara2023arXiv,Shilenko2023arXiv,Christiansson2023arXiv}. Shortly, the emergence of superconductivity in pressurized \LNO\ was confirmed by several groups~\cite{Hou2023arXiv,Zhang2023arXivYuan,Wang2024PRX}. Under ambient pressure, \LNO\ crystallizes in the orthorhombic $Amam$ structure. It is a paramagnetic metal with a phase transition at about 120~K~\cite{Taniguchi1995JPSJ,Kabayashi1996JPSJ,Wu2001PRB,Liu2023SCPMA} which has been suggested as a charge density wave (CDW)~\cite{Seo1996IC,Wu2001PRB,Liu2023SCPMA}. The application of pressure induces a structural transition from the $Amam$ to $Fmmm$ structure at about 10~GPa, and superconductivity with a maximum $T_{c}$ of 80~K emerges above 14~GPa. Theoretical work has underlined the crucial role of Ni-3$d$ orbitals and electronic correlations in the high-$T_{c}$ superconductivity in \LNO~\cite{Luo2023PRL,Zhang2023PRB,Yang2023arXiv,Lechermann2023arXiv,Gu2023arXiv,Shen2023arXiv,Sakakibara2023arXiv,Shilenko2023arXiv,Christiansson2023arXiv}. In this context, understanding the electronic correlations, charge dynamics and the role of Ni-3$d$ orbitals in \LNO\ is an important step towards the mechanism of the high-$T_{c}$ superconductivity and other instabilities in \LNO.

In this work, we investigate the optical properties of La$_{3}$Ni$_{2}$O$_{7}$. We find a substantial reduction of the electron's kinetic energy due to strong electronic correlations, which places La$_{3}$Ni$_{2}$O$_{7}$ near the Mott phase. Two Drude components are revealed in the low-frequency optical conductivity and ascribed to multiple bands formed by Ni-$d_{3z^2 - r^2}$ and Ni-$d_{x^2 - y^2}$ orbitals at the Fermi level. The transition at $T^{\ast}$ wipes out the Drude component with non-Fermi liquid behavior, and the one exhibiting Fermi-liquid behavior is not affected. These observations in conjunction with theoretical analysis point to the removal of the flat band formed by the Ni-$d_{3z^2 - r^2}$ orbital from the Fermi level upon the transition at $T^{\ast}$. Our experimental results shed light on the nature of the transition at $T^{\ast}$ and superconductivity in La$_{3}$Ni$_{2}$O$_{7}$.

%

%
%
\subsection*{Results}
\subsubsection*{Reflectivity and optical conductivity}
Figure~1a displays the temperature-dependent resistivity $\rho(T)$ of \LNO\ (red solid curve). While typical metallic behavior is realized from 300 down to 2~K, a kink occurs at about $T^{\ast} \simeq$~115~K, which has been observed in previous studies~\cite{Taniguchi1995JPSJ,Wu2001PRB,Liu2023SCPMA} and ascribed to a charge-density-wave (CDW) transition~\cite{Seo1996IC,Wu2001PRB,Liu2023SCPMA}. Figure~1b shows the far-infrared reflectivity $R(\omega)$ of \LNO\ at different temperatures; the spectra below 125~K are shifted down by 0.5 to better resolve the temperature dependence. $R(\omega)$ of \LNO\ approaches unity in the zero-frequency limit and increases with decreasing temperature in the far-infrared range, corroborating the metallic nature of the material. Below 125~K, the low-frequency $R(\omega)$ continues rising, whereas a suppression of $R(\omega)$ occurs between 400 and 1200~\icm. A comparison between the temperature dependence of $R(\omega)$ at 800~\icm\ (Fig.~1d) and $\rho(T)$ (Fig.~1a) links the suppression in $R(\omega)$ to the transition at $T^{\ast} \simeq$~115~K.

Figure~1c displays the real part of the optical conductivity $\sigma_{1}(\omega)$ for \LNO\ at different temperatures; the data below 125~K are shifted down by 1250~$\Omega^{-1}$cm$^{-1}$ to show the temperature dependence more clearly. The temperature dependence of $1/\sigma_{1}(\omega \rightarrow 0)$ (blue open circles in Fig.~1a) is compared with $\rho(T)$ (red solid curve in Fig.~1a) to verify the agreement between optical and transport measurements. A Drude peak is observed in the low-frequency $\sigma_{1}(\omega)$, which is the optical fingerprint of metals. As the temperature is lowered from 300~K to just above $T^{\ast}$, a progressively narrowing of the Drude response is observed. The narrowing of the Drude peak leads to a suppression of the high-frequency $\sigma_{1}(\omega)$ and an enhancement of the low-frequency $\sigma_{1}(\omega)$. Below $T^{\ast}$, the Drude peak is suppressed and the spectral weight [the area under $\sigma_{1}(\omega)$] is transferred to high frequency, resulting in a suppression of the low-frequency $\sigma_{1}(\omega)$ and an enhancement of the high-frequency $\sigma_{1}(\omega)$, which is opposite to the effect of the Drude peak narrowing above $T^{\ast}$. The inset of Fig.~1c shows a zoomed-in view of $\sigma_{1}(\omega)$ to highlight the spectral weight transfer caused by the transition at $T^{\ast}$ (the same data in a broader frequency range is shown in Supplementary Fig.~1). Figure~1e plots the value of $\sigma_{1}(\omega)$ at 1000~\icm\ as a function of temperature. The increase of $\sigma_{1}(1000~\text{cm}^{-1})$ occurs at $T^{\ast}$, indicating that the spectral weight transfer from low to high frequency is intimately related to the transition at $T^{\ast} \simeq 115$~K.

\subsubsection*{Drude-Lorentz analysis and theoretical calculations}
The measured $\sigma_{1}(\omega)$ of \LNO\ can be fitted to the Drude-Lorentz model,
%
%
\begin{equation}
\sigma_{1}(\omega) = \frac{2\pi}{Z_{0}} \left[\sum_{k} \frac{\omega^{2}_{p,k}}{\tau_{k}(\omega^{2}+\tau_{k}^{-2})}
   + \sum_{i} \frac{\gamma_{i} \omega^{2} \omega_{p,i}^{2}}{(\omega_{0,i}^{2} - \omega^{2})^{2} + \gamma_{i}^{2} \omega^{2}}\right],
\label{DLModel}
\end{equation}
where $Z_{0} \simeq 377$~$\Omega$ is the impedance of free space. The first term refers to a sum of Drude components which describe the optical response of free carriers or intraband transitions; each is characterized by a plasma frequency $\omega_{p}$ and a quasiparticle scattering rate $1/\tau$. The square of plasma frequency (Drude weight) $\omega_{p}^{2} = Z_{0}ne^{2}/2\pi m^{\ast}$, where $n$ and $m^{\ast}$ are the carrier concentration and effective mass, respectively. The second term represents a sum of Lorentzian oscillators which are used to model localized carriers or interband transitions. In the Lorentz term, $\omega_{0,i}$, $\gamma_{i}$, and $\omega_{p,i}$ are the resonance frequency (position), damping (line width), and plasma frequency (strength) of the $i$th excitation. \blue{Considering the multi-band nature of \LNO~\cite{Sun2023Nature,Luo2023PRL,Zhang2023PRB,Nakata2017PRB}, we use two Drude components to fit the data.} The cyan solid curve in Fig.~2a denotes the measured $\sigma_{1}(\omega)$ at 150~K, and the black dashed line through the data represents the fitting result, which is decomposed into two Drude components (red and blue shaded areas) and a series of Lorentz components (L1, green hatched area; L2, orange hatched area; L3, cyan hatched area; L4, purple hatched area; LH, grey hatched area). The inset of Fig.~2a shows the fitting result below 3000~\icm, highlighting the Drude components. The fitting parameters for all components are given in Supplementary Table~1, and the same parameters can also fit the imaginary part of the optical conductivity $\sigma_{2}(\omega)$ reasonably well (Supplementary Fig.~2).

In order to further understand the optical spectra of \LNO, we calculated the electronic band structure and $\sigma_{1}(\omega)$ for \LNO\ using first-principles density functional theory (DFT). In the calculated band structure (Fig.~2b), there are multiple bands crossing \EF: a flat hole-like band near the $\Gamma$ point and two broad electron-like bands near the $\Gamma$ and S (R) points. While the flat hole-like band and the broad electron-like band near the $\Gamma$ point arise from the Ni-$d_{3z^2 - r^2}$ (blue) and Ni-$d_{x^2 - y^2}$ (red) orbitals, respectively, the electron-like band near the S (R) point originates from mixed Ni-$d_{3z^2 - r^2}$ and Ni-$d_{x^2 - y^2}$ orbitals~\cite{Sun2023Nature}. These bands crossing \EF\ give rise to the two Drude components in $\sigma_{1}(\omega)$. The Lorentz components (L1, L2, L3, L4 and LH) are associated with interband electronic transitions. The red curve in Fig.~2c denotes the calculated $\sigma_{1}(\omega)$ without including the intraband contribution for \LNO\ which reproduces the features associated with interband transitions reasonably well. Nevertheless, the peak positions of the interband transitions in the experimental $\sigma_{1}(\omega)$ are shifted to slightly lower energies than those in the calculated $\sigma_{1}(\omega)$. This discrepancy between the measured and calculated $\sigma_{1}(\omega)$ is most likely related to electronic correlations~\cite{Si2009NP,Qazilbash2009NP,Geisler2024arXiv} which are not taken into account in DFT calculations. The blue curve in Fig.~2c represents the calculated $\sigma_{1}(\omega)$ including the intraband part for \LNO. It is noteworthy that the Drude profile in the measured $\sigma_{1}(\omega)$ (Fig.~2a) has significantly smaller weight than that in the calculated $\sigma_{1}(\omega)$ (blue curve in Fig.~2c), indicating strong electronic correlations in \LNO.

\subsubsection*{Electron's kinetic energy and electronic correlations}
The electronic correlations in a material can be obtained from the ratio $K_{\text{exp}}/K_{\text{band}}$~\cite{Millis2005PRB,Qazilbash2009NP,Schafgans2012PRL,Xu2020NC,Degiorgi2011NJP,Shao2020NP}, where $K_{\text{exp}}$ and $K_{\text{band}}$ refer to the experimental kinetic energy and the kinetic energy from band theory (DFT calculations), respectively. The kinetic energy of electrons is given by~\cite{Millis2005PRB,Qazilbash2009NP,Schafgans2012PRL}
%
%
\begin{equation}
K = \frac{2\hbar^{2} c_{0}}{\pi e^2}\int_{0}^{\omega_{c}}\sigma_{1}(\omega)\text{d}\omega,
\label{Kinetic}
\end{equation}
where $c_{0}$ is the $c$-axis lattice parameter, and $\omega_{c}$ is a cutoff frequency which should be high enough to cover the entire Drude component in $\sigma_{1}(\omega)$ but not so high as to include considerable contributions from interband transitions. For \LNO, due to the existence of low-energy interband transitions, as shown in Fig.~2a and 2c, the intraband and interband excitations strongly overlap with each other. In order to accurately determine $K_{\text{exp}}$, we subtract the interband contribution $\sigma^{\text{inter}}_{1}(\omega)$ from the total $\sigma^{\text{total}}_{1}(\omega)$ (Supplementary Fig.~3). The integral of $\sigma^{\text{total}}_{1}(\omega) - \sigma^{\text{inter}}_{1}(\omega)$ to $\omega_{c}$ = 3000~\icm\ yields $K_{\text{exp}}$ = 0.0258~eV. DFT calculations directly give $K_{\text{band}}$ = 1.17~eV (Supplementary Note~1). As a result, we get $K_{\text{exp}}/K_{\text{band}}$ = 0.022. Note that the choice of $\omega_{c}$ for $K_{\text{exp}}$ does not affect the value of $K_{\text{exp}}/K_{\text{band}}$, provided it covers the entire Drude response, because $K_{\text{exp}}/K_{\text{band}}$ quickly converges to $\sim$0.022 with increasing $\omega_{c}$ after the interband contribution is removed~\cite{Schafgans2012PRL} (Supplementary Fig.~4). Alternatively, we have that $K_{\text{exp}}/K_{\text{band}} = \omega_{p,\text{exp}}^{2}/\omega_{p,\text{cal}}^{2}$ (Supplementary Note~1). Here, $\omega_{p,\text{exp}} = \sqrt{\omega^2_{p,D1}+\omega^{2}_{p,D2}}$ = 3854~\icm\ (0.478~eV) is derived from the measured $\sigma_{1}(\omega)$ using the Drude-Lorentz fit; $\omega_{p,\text{cal}}$ = 3.22~eV is directly obtained from the DFT calculations (Supplementary Note~2). Consequently, we get $K_{\text{exp}}/K_{\text{band}}$ = 0.022, in good agreement with the value determined from Eq.~(2). Here, we would like to remark that the value of $K_{\text{exp}}/K_{\text{band}}$ does not change if a slight self-doping caused by oxygen deficiencies in \LNO\ is considered (Supplementary Note~3).

In Fig.~2d, we summarize $K_{\text{exp}}/K_{\text{band}}$ for \LNO\ (solid star) and some other representative materials (open symbols). For conventional metals such as Ag and Cu, $K_{\text{exp}}/K_{\text{band}}$ is close to unity, indicating negligible electronic correlations. In sharp contrast, the Mott insulator, e.g. the parent compound of the high-$T_{c}$ cuprate superconductor La$_{2}$CuO$_{4}$, has a vanishingly small $K_{\text{exp}}/K_{\text{band}}$, because the motion of electrons is impeded by strong on-site Coulomb repulsion, resulting in a substantial reduction of $K_{\text{exp}}$ compared to $K_{\text{band}}$. $K_{\text{exp}}/K_{\text{band}}$ in iron-based superconductors, for example LaOFeP and BaFe$_{2}$As$_{2}$, lie between conventional metals and Mott insulators, thus being categorized as moderately correlated materials~\cite{Qazilbash2009NP}. The value of $K_{\text{exp}}/K_{\text{band}}$ places \LNO\ in the proximity of the Mott insulator phase, closely resembling the doped cuprates~\cite{Millis2005PRB,Qazilbash2009NP}. This result suggests that in \LNO\ electronic correlations play an important role in the charge dynamics.

\subsubsection*{Temperature dependence of the charge dynamics}
By applying the Drude-Lorentz analysis to the measured $\sigma_{1}(\omega)$ at all temperatures (Supplementary Fig.~6), we extracted the temperature dependence of the Drude parameters. \blue{Figure~3a, b depict the temperature dependence of the weight for D1 ($\omega^{2}_{p,D1}$) and D2 ($\omega^{2}_{p,D2}$), respectively. While $\omega^{2}_{p,D1}$ exhibits no evident anomaly, $\omega^{2}_{p,D2}$ is suddenly suppressed below $T^{\ast} \simeq$~115~K and vanishes quickly with decreasing temperature. This implies that the Fermi surface is partially removed below the transition at $T^{\ast}$. Figure~3c, d plot the quasiparticle scattering rate of D1 ($1/\tau_{D1}$) and D2 ($1/\tau_{D2}$) as a function of temperature, respectively. It is worth noting that $1/\tau_{D1}$ follows a quadratic temperature dependence $1/\tau_{D1} \propto T^{2}$, i.e. Fermi-liquid behavior in a broad temperature range, whereas $1/\tau_{D2}$ varies linearly with temperature $1/\tau_{D2} \propto T$, which is the well-known non-Fermi-liquid behavior. These observations suggest that different Fermi surfaces in \LNO\ exhibit distinct electronic properties, and the portion exhibiting non-Fermi liquid behavior is removed due to the transition at $T^{\ast}$, leaving the portion characterized by Fermi liquid behavior to dominate the charge dynamics below $T^{\ast}$.} Furthermore, $\rho(T)$ exhibits non-Fermi liquid behavior above $T^{\ast}$ and Fermi liquid behavior below $T^{\ast}$ (Supplementary Fig.~7), in agreement with our optical results.

\subsubsection*{Energy scale of the spectral weight redistribution}
The transition at $T^{\ast}$ coincides with a suppression of $\omega^{2}_{p, D2}$ accompanied by a spectral weight transfer from low to high frequency in $\sigma_{1}(\omega)$. To find out the energy scale of the spectral weight redistribution, we examine the frequency and temperature dependence of the spectral weight defined as
%
%
\begin{equation}
S = \int_{0}^{\omega}\sigma_{1}(\omega)d\omega.
\label{SW}
\end{equation}
\blue{For a simple sharpening of the Drude response, the $S$ ratio, e.g. $S(200~\text{K})/S(300~\text{K})$ (the red curve in Fig.~4a) is higher than 1 in the low-frequency range and decreases smoothly towards 1 with increasing frequency.} By contrast, $S(5~\text{K})/S(125~\text{K})$ (the blue curve in Fig.~4a) exhibits completely different behavior, which gives the direction and energy scale of the spectral weight transfer associated with the transition at $T^{\ast}$. In the low-frequency limit, the large value of $S(5~\text{K})/S(125~\text{K})$ results from the narrowing of the Drude peak at low temperatures. With increasing frequency, $S(5~\text{K})/S(125~\text{K})$ decreases steeply and reaches a minimum smaller than 1 at about 600~\icm. This indicates that the spectral weight below 600~\icm\ is significantly suppressed at 5~K. Note that the spectral weight below 600~\icm\ is mainly contributed by the Drude components (see the inset of Fig.~2a), so the sharp decrease of $S(5~\text{K})/S(125~\text{K})$ is consistent with the suppression of $\omega^{2}_{p, D2}$ below $T^{\ast}$. As the frequency further increases, $S(5~\text{K})/S(125~\text{K})$ rises monotonically and reaches unity (black dashed line) at about 6500~\icm. This behavior suggests that the lost low-frequency spectral weight is retrieved in a very broad frequency range up to 6500~\icm. Figure~4c-e plot $S(T)/S(300~\text{K})$ for different cutoff frequencies as a function of temperature. For low cutoff frequencies, such as $\omega_{c}$ = 600~\icm\ (Fig.~4c) and $\omega_{c}$ = 1000~\icm\ (Fig.~4d), $S(T)/S(300~\text{K})$ increases upon cooling from 300~K, which is caused by the narrowing of the Drude response in $\sigma_{1}(\omega)$ and strong electronic correlation effect. In Mott systems, the electron's kinetic energy (Drude weight) increases with decreasing temperature~\cite{Kotliar2004PT,Basov2011RMP}. This behavior is evident particularly for temperatures above $T^{\ast}$, suggesting that the proximity of \LNO\ to a Mott phase may be applicable primarily to the temperature range exhibiting non-Fermi-liquid behavior. Below $T^{\ast}$, $S(T)/S(300~\text{K})$ decreases, in good agreement with the suppression of $\omega^{2}_{p,D2}$. $S(T)/S(300~\text{K})$ for $\omega_{c}$ = 6500~\icm\ (Fig.~4e) is essentially temperature independent, indicating that the removed low-frequency spectral weight due to the transition at $T^{\ast}$ is fully recovered at 6500~\icm. \blue{Temperature-dependent ellipsometry would be a more accurate technique to quantify the spectral weight redistribution in the high-frequency range.}

%
%
\subsection*{Discussion}
Our optical results show that \LNO\ features strong electronic correlations which substantially reduce the electron's kinetic energy and place the material in the proximity of the Mott phase. The electronic correlation strength in \LNO\ is comparable to that in doped cuprates~\cite{Millis2005PRB,Qazilbash2009NP}, but much stronger than that in iron-based superconductors~\cite{Millis2005PRB,Qazilbash2009NP,Degiorgi2011NJP}. Interestingly, the maximum $T_{c}$ in \LNO\ is also comparable to that in cuprates but higher than that in iron-based superconductors. This coincidence is likely to hint that the high-$T_{c}$ superconductivity in \LNO\ is intimately related to electronic correlations. In cuprate systems, the Cu-$d_{x^2 - y^2}$ orbital plays a significant role in the strong electronic correlations and superconductivity~\cite{Imada1998RMP}, whereas \LNO\ is a multi-orbital system with both the Ni-$e_{g}$ (Ni-$d_{3z^2 - r^2}$ and Ni-$d_{x^2 - y^2}$) orbitals crossing \EF. Recent theoretical calculations have revealed strong electronic correlations in \LNO\ particularly for the Ni-$d_{3z^2 - r^2}$ orbital~\cite{Lechermann2023arXiv,Shilenko2023arXiv} and emphasized their important role in promoting a superconducting instability~\cite{Lechermann2023arXiv}. Given the mixture of both Ni-$e_{g}$ orbitals at specific $k$-points, the Ni-$d_{x^2 - y^2}$ orbital is likely to have considerable influence on the electronic correlations and superconductivity as well.

The temperature evolution of the Drude parameters has demonstrated that different Fermi surfaces in \LNO\ exhibit distinct electronic properties, and the portion exhibiting non-Fermi liquid behavior is removed below the phase transition at $T^{\ast}$, leaving the portion with Fermi liquid behavior to dominate the charge dynamics below $T^{\ast}$. Theoretical work~\cite{Lechermann2023arXiv,Shilenko2023arXiv} has revealed remarkable orbital differentiation in \LNO\ with the Ni-$d_{3z^2 - r^2}$ orbital being more strongly correlated, thus inducing spin fluctuations and non-Fermi liquid behavior. A recent angle-resolved photoemission spectroscopy (ARPES) study has also found orbital-dependent electronic correlations in \LNO\ with the Ni-$d_{3z^2 - r^2}$ derived flat band $\gamma$ showing much stronger electronic correlations than the Ni-$d_{x^2 - y^2}$ derived bands $\alpha$ and $\beta$~\cite{Yang2023arXivARPES}. The combination of these facts and our optical results implies that the removed Fermi surface below $T^{\ast}$ corresponds to the flat hole-like band arising from the Ni-$d_{3z^2 - r^2}$ orbital, and the remaining Fermi surface consists of the broad electron-like bands near the $\Gamma$ and S (R) points. The remaining Fermi surface is dominated by the Ni-$d_{x^2 - y^2}$ orbital which has a quarter filling in \LNO~\cite{Nakata2017PRB,Sun2023Nature}, corresponding to the heavily overdoped regime in the phase diagram for hole-doped cuprates, where a Fermi liquid is found despite the strong electronic correlations~\cite{Keimer2015Nature,Imada1998RMP}. This fact may account for the observed Fermi liquid behavior for D1. ARPES data at 18~K has shown that while the Ni-$d_{x^2 - y^2}$ derived bands ($\alpha$ and $\beta$) cross \EF, the Ni-$d_{3z^2 - r^2}$ derived band ($\gamma$) is below \EF~\cite{Yang2023arXivARPES}. This result seems consistent with our analysis. However, temperature-dependent ARPES studies are required to track the evolution of the Ni-$d_{3z^2 - r^2}$ derived band across the transition at $T^{\ast}$.

The Fermi surface reduction and spectral weight redistribution below $T^{\ast}$ may be associated with a density-wave gap whose characteristic optical response is a suppression of the Drude weight in $\sigma_{1}(\omega)$ alongside a spectral weight transfer from low to high frequencies~\cite{Hu2008PRL,Zhou2021PRB,Homes2012PRB,Tediosi2009PRB}. Many theoretical and experimental studies~\cite{Seo1996IC,Shilenko2023arXiv,Shen2023arXiv,Yang2023arXiv,Chen2023arXiv,Wu2001PRB,Liu2023SCPMA} have suggested charge and spin density wave instabilities in \LNO. The trilayer R-P phase La$_{4}$Ni$_{3}$O$_{10}$ exhibits a similar transition at about 140~K, which has been demonstrated to be intertwined incommensurate charge and spin density waves~\cite{Zhang2020NC}. Moreover, ARPES on La$_{4}$Ni$_{3}$O$_{10}$ has revealed a pseudogap opening in the band of Ni-$d_{3z^2 - r^2}$ orbital character below the transition~\cite{Li2017NC}. Similar intertwined incommensurate charge and spin density waves accompanied by a gap opening may also occur in \LNO, accounting for the Fermi surface reduction and spectral weight redistribution. The gap value $\Delta$ can be extracted from the difference optical conductivity~\cite{Uykur2022NPJQM,Zhou2023PRBCVNS,Zhou2023PRBAVS} $\Delta \sigma_{1}(\omega) = \sigma_{1}^{T < T^{\ast}}(\omega) - \sigma_{1}^{N}(\omega)$, where $\sigma_{1}^{T < T^{\ast}}(\omega)$ and $\sigma_{1}^{N}(\omega)$ denote $\sigma_{1}(\omega)$ at $T < T^{\ast}$ and $\sigma_{1}(\omega)$ in the normal state, respectively. Here for \LNO, $\sigma_{1}(\omega)$ at 125~K is used as $\sigma_{1}^{N}(\omega)$. Figure~4b shows $\Delta \sigma_{1}(\omega)$ at 5~K, in which the zero-crossing point, as indicated by the black arrow, corresponds to $2\Delta$ = 100.5~meV, resulting in a ratio of $2\Delta/k_{\text{B}}T^{\ast}$ = 10.14 that is much larger than the weak-coupling BCS value 3.52. ARPES measurements have shown that the Ni-$d_{3z^2 - r^2}$ derived band ($\gamma$) is about 50~meV below the Fermi level at 18~K~\cite{Yang2023arXivARPES}. This energy scale is in excellent agreement with the value of $\Delta$ = 50.25~meV we extracted from $\Delta \sigma_{1}(\omega)$. In addition, the temperature dependence of $\Delta$ (red open circles in Fig.~4f) deviates from the BCS mean-field behavior (blue solid line in Fig.~4f) near the transition temperature $T^{\ast}$. These observations are clearly at odds with the conventional description of density waves~\cite{Peierls1955,Frohlich1954PRSLA,Gruner1988RMP,Gruner1994RMP}. The large $2\Delta/k_{\text{B}}T^{\ast}$ ratio in \LNO\ suggests that the coherence length of the density wave order is small~\cite{McMillan1977PRB}. With a short coherence length, strong critical fluctuations are expected near the transition temperature, which is likely to be responsible for the discrepancy between the temperature dependence of $\Delta$ and the BCS mean-field behavior near $T^{\ast}$~\cite{McMillan1977PRB}. \blue{On the other hand, since theoretical calculations (Fig.~2b and Refs.~\cite{Sun2023Nature,LaBollita2023arXiv,Nakata2017PRB}) have shown that the Ni-$d_{3z^2 - r^2}$ derived flat band near $\Gamma$ is in close proximity to \EF, which is subject to strong temperature effect, one may argue that a simple temperature-induced Fermi energy shift can also account for the Fermi surface reduction and spectral weight redistribution. Here, we would like to point out that for a simple temperature-induced Fermi energy shift, the Drude weight is expected to decrease continuously as the temperature is lowered~\cite{Xu2016PRBTaAs}. However, in \LNO, the Drude weight does not show a continuous decrease upon cooling, but is abruptly suppressed below the transition, at odds with the temperature-induced Fermi level shift.}

It is instructive to compare \LNO\ with the infinite-layer nickelate system such as LaNiO$_{2}$. Both \LNO\ and LaNiO$_{2}$ are multi-orbital electronic systems with dominant Ni-$3d$ orbitals, electronic correlations and a tendency towards orbital differentiation~\cite{Wang2020PRB,Kang2021PRL}. However, while LaNiO$_{2}$ with $K_{\text{DMFT}}/K_{\text{DFT}}$ = 0.5--0.6 is far from a Mott phase but close to a Hund's metal~\cite{Wang2020PRB,Kang2021PRL}, \LNO\ is in proximity to the Mott regime ($K_{\text{exp}}/K_{\text{band}}$ = 0.022). Moreover, in \LNO\ the Ni-$d_{3z^{2}-r^{2}}$ orbital exhibits stronger electronic correlations than the Ni-$d_{x^{2}-y^{2}}$ orbital, which contrasts sharply with the situation of LaNiO$_{2}$ in which the Ni-$d_{x^{2}-y^{2}}$ orbital is more strongly correlated than the Ni-$d_{3z^{2}-r^{2}}$ orbital~\cite{Wang2020PRB,Kang2021PRL}. These differences may have significant influence on the ground states of the two systems.

Finally, we would like to underline that \LNO\ only exhibits superconductivity under pressure~\cite{Sun2023Nature,Hou2023arXiv,Zhang2023arXivYuan,Wang2024PRX}, and a structural phase transition occurs at around 10~GPa in tandem with superconductivity. Therefore, our optical study at ambient pressure can not provide direct information about the superconductivity in this system. Nevertheless, the pressure-induced structural transition from the $Amam$ to $Fmmm$ phase does not change the main characteristics of the electronic structure, such as the dominant Ni-$d_{3z^2 - r^2}$ and Ni-$d_{x^2 - y^2}$ orbitals near the Fermi level as well as electronic bonding and anti-bonding bands of the Ni-$d_{3z^{2}-r^{2}}$ orbital, which are believed to be important to superconductivity~\cite{Sun2023Nature,LaBollita2023arXiv}. In addition, the superconductivity in \LNO\ is achieved by suppressing the density-wave-like transition at $T^{\ast}$ in the ambient-pressure phase, which is probably a competing order~\cite{Zhang2023arXivYuan,Wang2024PRX}. In this context, studying the spectroscopic properties of the dominant orbitals and the possible competing order in the ambient-pressure phase could also provide important information for understanding the superconductivity in pressurized \LNO. \blue{Considering the proximity of \LNO\ to the Mott state, the density-wave-like transition in \LNO\ may be driven by strong electronic correlations. High pressures may weaken the electronic correlations and suppress the density-wave-like order, allowing superconductivity to emerge.}

To summarize, our optical study reveals strong electronic correlations in La$_{3}$Ni$_{2}$O$_{7}$ which give rise to a substantial reduction of the electron's kinetic energy and place this compound near the Mott insulator phase. Multiple bands dominated by Ni-$d_{3z^2 - r^2}$ and Ni-$d_{x^2 - y^2}$ orbitals cross the Fermi level, resulting in the presence of two Drude components in the low-frequency optical conductivity. Below the transition at $T^{\ast}$, the Drude component exhibiting non-Fermi liquid behavior is removed, leaving the one with Fermi-liquid behavior to dominate the charge dynamics. These observations in combination with theoretical calculations suggest that the Fermi surface associated with the strongly correlated flat band derived from the Ni-$d_{3z^2 - r^2}$ orbital is removed. Our experimental results provide key information for understanding the nature of the transition at $T^{\ast}$ and superconductivity in La$_{3}$Ni$_{2}$O$_{7}$.



%
%
\subsection*{Methods}
\begin{trivlist}
\item[]\textbf{Single crystal growth.} High-quality single crystals of \LNO\ were grown by a vertical optical-image floating zone furnace with an oxygen pressure of 15~bar and a 5~kW Xenon arc lamp (100-bar Model HKZ, SciDreGmbH, Dresden).

\item[]\textbf{Optical measurements and Kramers-Kronig analysis.} The near-normal-incidence \emph{ab}-plane reflectivity $R(\omega)$ at ambient pressure was measured in the frequency range of 30--50\,000~\icm\ using a Bruker Vertex 80v Fourier transform infrared spectroscopy (FTIR). An \emph{in situ} gold/silver evaporation technique~\cite{Homes1993} was adopted. The real part of the optical conductivity $\sigma_{1}(\omega)$ was determined via a Kramers-Kronig analysis of the measured $R(\omega)$ for \LNO~\cite{Dressel2002,Tanner2019}. Below the lowest measured frequency (30~\icm), a Hagen-Rubens ($R = 1 - A\sqrt{\omega}$) form was used for the low-frequency extrapolation. Above the highest measured frequency, we assumed a constant reflectivity up to 12.5~eV, followed by a free-electron ($\omega^{-4}$) response.

\item[]\textbf{DFT calculations.} The density functional theory (DFT) calculations were performed using the all-electron, full-potential WIEN2K code with the augmented plane-wave plus local orbital (APW+lo) basis set~\cite{Blaha2001} and the Perdew-Burke-Ernzerhof (PBE) exchange functional~\cite{Perdew1996PRL}. A total number of $12 \times 12 \times 12$ k-points in the reduced first brillouin zone was used for the self-consistency cycle. The optical properties were calculated with a k-points mesh of $33 \times 33 \times 33$ in the first Brillouin zone to ensure convergency. The broadening factor (scattering rate) employed in computing $\sigma_{1}(\omega)$ is 0.025~eV, which corresponds to the average of the Drude width $(1/\tau_{D1}+1/\tau_{D2})/2$ = 205~\icm\ ($\sim$0.025~eV) in the experimental $\sigma_{1}(\omega)$ at 150~K. All calculations were performed using the experimental structure under ambient pressure~\cite{Ling2000JSSC}. In addition, as the optical spectra were measured in the $ab$ planes and \LNO\ crystallizes in the orthorhombic $Amam$ structure, the calculated optical conductivity was given by $\sigma_{1}(\omega) = (\sigma_{xx}(\omega) + \sigma_{yy}(\omega))/2$.
\end{trivlist}

\subsection*{Data availability}
All data that support the findings of this study are available from the corresponding authors upon request.

\subsection*{Code availability}
The computer code used for data analysis is available upon request from the corresponding author.

\subsection*{Acknowledgements}
We thank C.~C.~Homes, J.~Schmalian, Qianghua~Wang, Bing~Xu, Huan~Yang, Run~Yang, Dao-Xin~Yao, Shunli~Yu, Peng~Zhang and Guang-Ming~Zhang for helpful discussions. Work at Nanjing University was supported by the National Key R\&D Program of China (Grants No. 2022YFA1403201 and No. 2022YFA1403000), the National Natural Science Foundation of China (Grants No. 12174180, No. 12274207, No. 12204231, and No. 12061131001), and Jiangsu shuangchuang program. Work at Sun Yat-Sen University was supported by the National Natural Science Foundation of China (Grant No. 12174454), Guangdong Basic and Applied Basic Research Funds (Grant no. 2021B1515120015), and Guangdong Provincial Key Laboratory of Magnetoelectric Physics and Devices (Grant no. 2022B1212010008).

\subsection*{Author contributions}
Z.L. performed the optical measurements with the assistance of J.H., X.Z. and Y.D.; M.H. and M.W. synthesized the crystals; Q.L. and Y.L. characterized the samples; J.L. and Y.L. performed DFT calculations; Z.L., Y.D. and H.H.W. analyzed the data and wrote the manuscript; all authors made comments on the manuscript.

\subsection*{Competing interests}
The authors declare no competing interests.

\subsection*{Additional information}

\begin{trivlist}
\item[]\textbf{Extended data} accompanies this paper at ...

\item[]\textbf{Supplementary information} accompanies this paper at ...

\item[]\textbf{Correspondence and requests for materials} should be addressed to\\
Yaomin~Dai, Yi~Lu, Meng~Wang or Hai-Hu~Wen.

\item[]\textbf{Reprints and permissions information} is available online.
\end{trivlist}


%
%
\clearpage
\centerline{\includegraphics[width=\columnwidth]{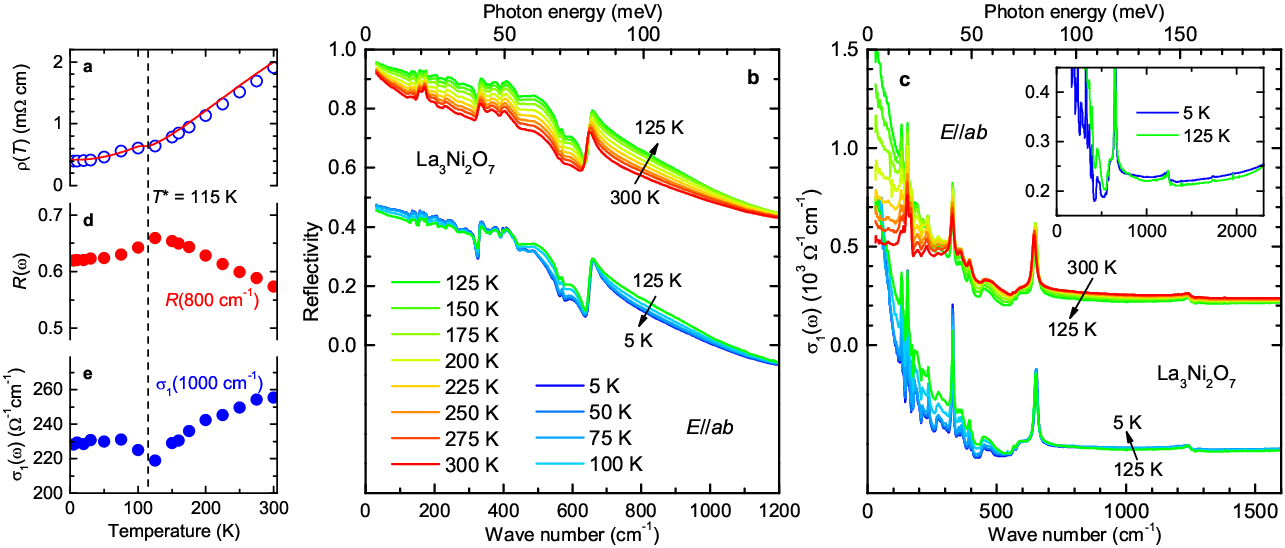}}%
\vspace*{0.8cm}
\noindent{\textsf{\textbf{Figure 1 $|$ Reflectivity and optical conductivity of La$_{3}$Ni$_{2}$O$_{7}$.} \textbf{a} The temperature-dependent resistivity $\rho(T)$ obtained from transport measurements (red solid curve) and that from optical measurements (blue open circles). \textbf{b} The far-infrared \emph{ab}-plane reflectivity $R(\omega)$ of La$_{3}$Ni$_{2}$O$_{7}$ at different temperatures. The spectra below 125~K are shifted down by 0.5 to show the temperature dependence more clearly. \textbf{c} The optical conductivity $\sigma_{1}(\omega)$ of La$_{3}$Ni$_{2}$O$_{7}$ at different temperatures in the far-infrared range. The curves below 125~K are shifted down by 1250~$\Omega^{-1}$cm$^{-1}$. The inset shows an enlarge view of $\sigma_{1}(\omega)$ to highlight the spectral weight transfer induced by the transition at $T^{\ast}$. \textbf{d}, \textbf{e} show the temperature dependence of $R(\omega)$ at 800~\icm\ and $\sigma_{1}(\omega)$ at 1000~\icm, respectively. Both exhibit a clear anomaly at $T^{\ast}$ = 115~K.}}

%
%
\clearpage
\centerline{\includegraphics[width=0.8\columnwidth]{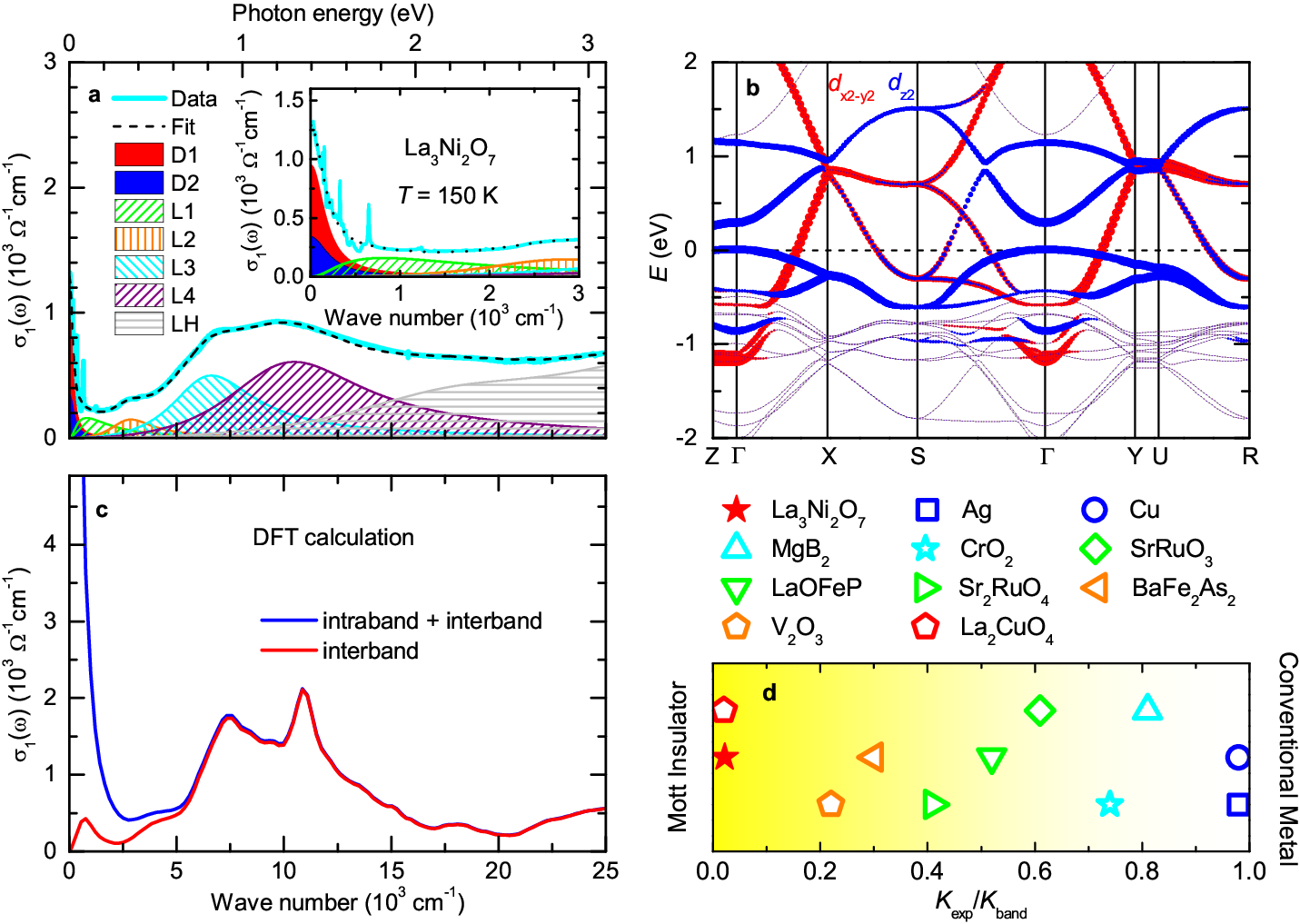}}%
\vspace*{0.8cm}
\noindent{\textsf{\textbf{Figure 2 $|$ Drude-Lorentz fit and theoretical calculations.} \textbf{a} The measured $\sigma_{1}(\omega)$ of \LNO\ at 150~K (cyan solid curve) and the Drude-Lorentz fitting result (black dashed line). The fitting curve is decomposed into two Drude components (red and blue shaded areas) and a series of Lorentz components L1 (green hatched area), L2 (orange hatched area), L3 (cyan hatched area), L4 (purple hatched area) and LH (grey hatched area). The inset shows an enlarged view of the fitting result in the low-frequency range. \textbf{b} The calculated electronic band structure for \LNO. \textbf{c} The calculated $\sigma_{1}(\omega)$ of \LNO. \textbf{d} $K_{\text{exp}}/K_{\text{band}}$ for \LNO\ (solid star) and some other representative materials. $K_{\text{exp}}/K_{\text{band}}$ for other materials are taken from Ref.~\cite{Qazilbash2009NP} and the references cited therein.}}

%
%
\clearpage
\centerline{\includegraphics[width=0.7\columnwidth]{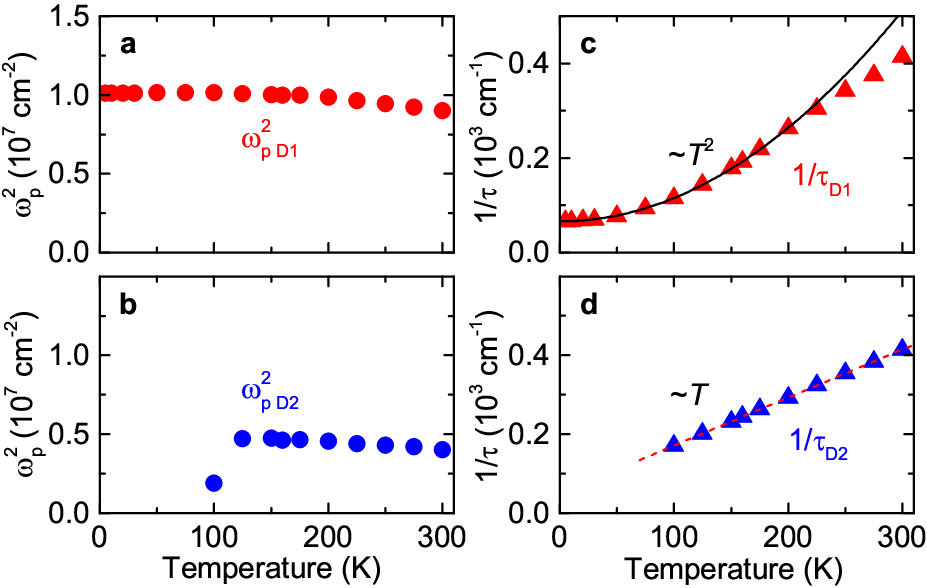}}%
\vspace*{0.8cm}
\noindent{\textsf{\textbf{Figure 3 $|$ Temperature dependence of the Drude parameters.} The temperature dependence of the weight for D1 (\textbf{a}) and D2 (\textbf{b}). \textbf{c} The quasiparticle scattering rate for D1 as a function of temperature. The solid line denotes a quadratic temperature dependence. \textbf{d} The temperature dependence of the quasiparticle scattering rate for D2. The dashed line is a linear fit.}}

%
%
\clearpage
\centerline{\includegraphics[width=0.7\columnwidth]{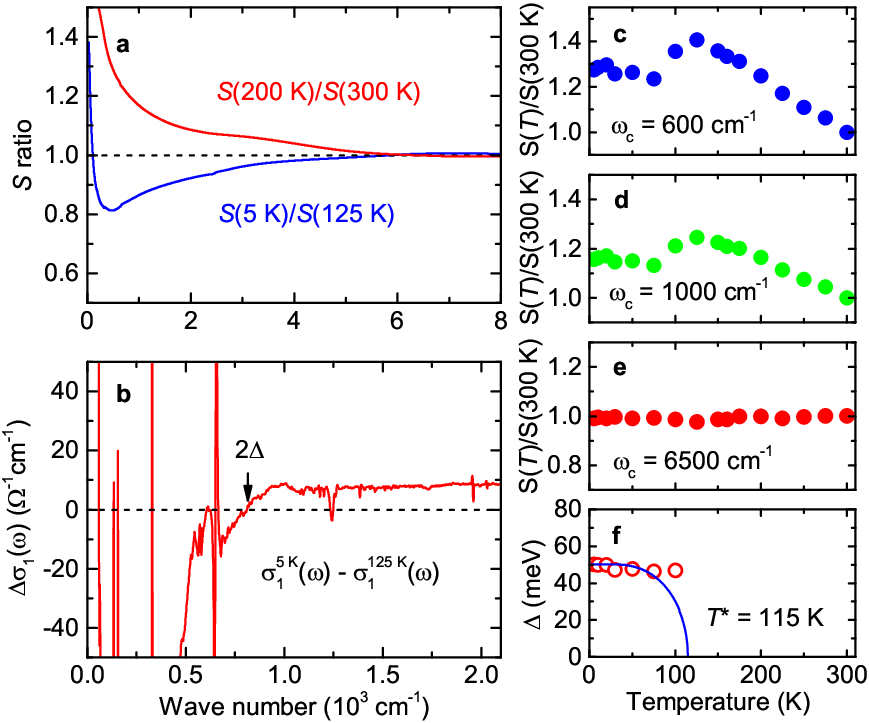}}%
\vspace*{0.8cm}
\noindent{\textsf{\textbf{Figure 4 $|$ Energy scale of the gap and spectral weight redistribution.} \textbf{a} The frequency-dependent spectral weight ratio $S(200~\text{K})/S(300~\text{K})$ (red curve) and $S(5~\text{K})/S(125~\text{K})$ (blue curve). \textbf{b} The difference optical conductivity $\Delta \sigma_{1}(\omega)$ at 5~K. The arrow indicates the zero-crossing point in $\Delta \sigma_{1}(\omega)$ which corresponds to the gap energy 2$\Delta$. \textbf{c}--\textbf{e} The evolution of the spectral weight as a function of temperature for cutoff frequencies $\omega_{c}$ = 600~\icm, $\omega_{c}$ = 1000~\icm\ and $\omega_{c}$ = 6500~\icm. \textbf{f} The evolution of $\Delta$ with temperature (red open circles). The blue solid line denotes the mean-field behavior.}}


\end{document}